\documentclass[12pt]{iopart}
\usepackage{amssymb,latexsym,epsfig}
\newcommand{\beq}{\begin{equation}}
\newcommand{\eeq}{\end{equation}}
\newcommand{\bqa}{\begin{eqnarray}}
\newcommand{\eqa}{\end{eqnarray}}
\newcommand{\fr}{\frac}

\begin{document}
\title{Strong curvature singularities in quasispherical asymptotically de Sitter dust collapse}
\author{S\' ergio M. C. V. Gon\c calves}
\address{Theoretical Astrophysics, California Institute of Technology, Pasadena, California 91125, U.S.A.}
\date{\today}
\begin{abstract}
We study the occurrence, visibility, and curvature strength of singularities in dust-containing Szekeres spacetimes (which possess no Killing vectors) with a positive cosmological constant. We find that such singularities can be locally naked, Tipler strong, and develop from a non-zero-measure set of regular initial data. When examined along timelike geodesics, the singularity's curvature strength is found to be independent of the initial data.
\end{abstract}
\pacs{0420D, 0420J, 0470B}
\maketitle

\section{Introduction}

It has recently been shown that a central, locally naked, Tipler strong singularity develops in the gravitational collapse of inhomogeneous spherical dust with a positive cosmological constant \cite{goncalves00,deshingkaretal00}. Those results are a generalization to asymptotically de Sitter spacetimes of the well-known asymptotically flat spherical dust---Lema\^{\i}tre-Tolman-Bondi (LTB)---collapse models. The exact solvability of such models enables a detailed analysis of gravitational collapse, and formation and structure of singularities, which shows that, from regular initial data, a Tipler strong curvature singularity always develops, which can be locally or globally naked, depending on the initial data \cite{tbauthors}. Although almost a paradigm for gravitational collapse, LTB models rely on the simplifying assumptions of irrotacional dust and spherical symmetry. Within spherical symmetry, various generalizations of the matter model have been considered \cite{ltbg3,ltbg4,ltbg5,ltbg6,ltbg7}, with similar results: a central curvature singularity forms, and its visibility (local vs. global) depends on the initial data.

Comparatively, much less is known about singularity formation and structure in non-spherical collapse. Thorne's seminal analysis of cylindrical collapse \cite{thorne65} led him to formulate the {\em hoop conjecture}, which essentially states that horizons form if and only when the gravitational mass of the system is confined to a maximum ``radius'' in every direction \cite{thorne72}. Subsequent numerical analyses of prolate and oblate collapse by Shapiro and Teukolsky \cite{shapiro&teukolsky}, and of gravitational radiation emission in aspherical collapse by Nakamura, Shibata, and Nakao \cite{nakamura&shibata&nakao}, were unable to refute Thorne's conjecture. A perturbative analysis of LTB collapse by Hirada, Iguchi, and Nakao \cite{hin1,hin2,hin3}, showed that linear non-spherical matter, metric, and matter coupled to metric perturbations remain bounded in the limit of approach to the Cauchy horizon. The very similar structure of the central singularity in LTB and Tolman-Bondi-de Sitter (TBdS) models \cite{goncalves00,barveetal99}, strongly suggests that analogous marginal stability properties also hold for the latter case. The quasispherical dust collapse models, given by the Szekeres metric \cite{szekeres75a}, were analyzed by Szekeres himself \cite{szekeres75b}, Joshi and Kr\' olak \cite{joshi&krolak96}, and by Deshingkar, Joshi, and Jhingan \cite{deshingkar&jhingan&joshi98}. As in the spherical LTB case, a ``central'' singularity forms, which is Tipler strong, and can be locally or globally naked, depending on the initial data.

Recent measurements of type Ia supernovae \cite{riess98,perlmutter99} and peculiar motion of low redshift galaxies \cite{zehani&dekel99}, appear to suggest the existence of a positive cosmological constant. The possibility of non-zero asymptotic (constant) curvature constitutes an obvious motivation for the study of gravitational collapse with a cosmological constant \cite{goncalves00,deshingkaretal00,markovic&shapiro00}. In this paper, we address one aspect of this problem by studying the formation, visibility, and curvature strength of singularities in quasispherical inhomogeneous spacetimes with a positive cosmological constant, thereby generalizing the results of \cite{goncalves00} to the quasispherical case, and those of \cite{szekeres75b,joshi&krolak96,deshingkar&jhingan&joshi98} to the asymptotically de Sitter case.

The general metric for Szekeres spacetimes with a positive cosmological constant was obtained in closed form by Barrow and Stein-Schabes \cite{barrow&steinschabes84}, in the context of the ``cosmic no hair theorem'', but this solution also provides a valuable test-bed metric for the analysis of singularity formation and structure in non-spherical gravitational collapse. One must remark, however, that such a departure from spherical symmetry, far from being arbitrary, is very well-defined---whilst the four-dimensional spacetime metric does not admit any Killing vectors, it does possess an invariant family of spherical two-surfaces, hence the name {\em quasispherical}. Accordingly, the results presented here provide a useful, but somewhat limited, insight into non-perturbative departures from spherical symmetry. Note also that, Szekeres spacetimes can be matched to a Schwarzschild (or Schwarzschild-de Sitter, if $\Lambda>0$) spacetime, and thus cannot contain gravitational waves \cite{bonnor76}.

As we shall see below, as in the Szekeres and TBdS cases, for quasispherical asymptotically de Sitter dust collapse, from regular initial data a central singularity develops, which is Tipler strong, and locally visible (for a large class of initial data). Our results suggest that neither lack of asymptotic flatness, nor ``mild'' deviations (in a sense to be precisely defined below) from spherical symmetry, play an important role in the formation and nature of singularities in gravitational collapse.

The paper is organized as follows: Szekeres spacetimes with a cosmological constant are outlined in section 2. Section 3 discusses the existence and visibility of the singularity, along null and timelike directions. The curvature strength of the singularity is analyzed in section 4. Section 5 concludes with a discussion and summary.

Geometrized units, in which $G=c=1$, are used throughout.

\section{Szekeres spacetimes with a cosmological constant}

In this section we present the relevant equations for Szekeres metrics \cite{szekeres75a,szekeres75b} with a positive cosmological constant \cite{barrow&steinschabes84}. The stress-energy tensor is 
\beq
T_{ab}=(p+\rho)u_{a}u_{b}-\left(p-\fr{\Lambda}{8\pi}\right)g_{ab}, \;\;\; u_{a}=\delta^{t}_{a},
\eeq
where $u_{a}$, $p$ and $\rho$ are the four-velocity, pressure and fluid density, respectively. The metric can be written in normal Gaussian coordinates:
\beq
ds^{2}=-dt^{2}+e^{2\alpha}dr^{2}+R^{2}e^{2\nu}(dx^{2}+dy^{2}), \label{szm}
\eeq
where $\alpha=\alpha(t,r,x,y)$, $R=R(t,r)$, $\nu=\nu(r,x,y)$, and 
\bqa
e^{\alpha(t,r,x,y)}&=&[R'(t,r)+R(t,r)\nu'(r,x,y)][1+k(r)]^{-\fr{1}{2}}, \label{alph} \\
e^{-\nu(r,x,y)}&=&c_{1}(x^{2}+y^{2})+2(c_{2}x+c_{3}y)+c_{4}, \label{nu} \\
k(r)&=&4(c_{1}c_{4}-c_{2}^{2}-c_{3}^{2})-1, \label{kay}
\eqa
where $'\equiv\partial_{r}$, $\dot{\,}\equiv\partial_{t}$, and the real-valued functions $c_{i}=c_{i}(r)$, $i=1...4$, are to be specified within each Szekeres class. The Szekeres metrics (with or without a cosmological constant) can be divided into two classes, depending on whether $(Re^{\nu})'$ vanishes or not \cite{szekeres75a,barrow&steinschabes84}. The class defined by $(Re^{\nu})'=0$ contains shell-crossing singularities \cite{yodzis&seifert&hagen73,goncalves01b}, which are physically ``mild''---they are gravitationally weak \cite{newman86} and geodesically complete \cite{papapetrou&hamoui67}---and hence will not be considered here. We shall focus on the class for which $(Re^{\nu})'\neq0$, which admits shell-focusing curvature singularities, but no shell-crossings.

The function $R(t,r)$ obeys the evolution equation
\beq
\dot{R}^{2}=\fr{2M(r)}{R}-k(r)+\fr{1}{3}\Lambda R^{2}, \label{szd}
\eeq
where $M(r)$ is a real-valued free function. Introducing an auxiliary variable $\eta$, defined by
\beq
\sqrt{R}d\eta=dt,
\eeq
Equation (\ref{szd}) reduces to
\beq
\fr{d\eta}{d\theta}=\pm\sqrt{\fr{6}{\Lambda p}}(\gamma\pm\cosh\theta)^{-\fr{1}{2}}, \label{dedt}
\eeq
which has solutions of the form
\beq
\eta=\eta_{0}\pm2\sqrt{\fr{6}{\Lambda p}}(\gamma+1)^{-\fr{1}{2}}F(\psi,a),
\eeq
where $\eta_{0}=\eta(\theta=0)$, and $F(\psi,a)$ is an elliptic function of the third kind \cite{abramovitz&stegun64}, with $\psi\equiv\sin^{-1}\Psi$, where
\bqa
\Psi&\equiv&\tanh(\theta/2) \;\;\;\;\;\; [\mbox{``+'' in} \;(\ref{dedt})], \\
\Psi&\equiv&\sqrt{(\gamma-\cosh\theta)/(\gamma-1)} \;\;\;\;\;\; [\mbox{``-'' in} \;(\ref{dedt})],
\eqa
and
\beq
4\left(R^{2}+R\chi-\fr{3M}{\Lambda\chi}\right)=\left(\chi^{2}-\fr{12M}{\Lambda\chi}\right)\sinh^{2}\theta, 
\eeq
where
\bqa
\gamma&\equiv&-3\fr{\chi}{p}, \\
\chi&=&\begin{array}{lll}
\left\{\begin {array}{lll}
-|4k/\Lambda|^{\fr{1}{2}}\cosh\Theta & , & k\geq0, \\
-|-4k/\Lambda|^{\fr{1}{2}}\sinh\Theta & , & k\leq0,
\end{array}
\right.
\end{array} \\
\Theta&\equiv&\fr{1}{3}\cosh^{-1}\left(\fr{3M}{2}\sqrt{\left|\fr{-\Lambda}{k^{3}}\right|}\right), \\
p&\equiv&\sqrt{\chi^{2}+\fr{12M}{\Lambda\chi}}, \\
a&\equiv&\sqrt{(\gamma-1)/(\gamma+1)}.
\eqa

For the particular case of dust ($p=0$), with gravitationally unbound matter distributions ($k=0$), the solution reduces to the simple algebraic form:
\beq
R(t,r)=\left(\fr{3M}{\Lambda}\right)^{\fr{1}{3}}\sinh^{\fr{2}{3}}\left\{\sqrt{\fr{3\Lambda}{4}}[t_{\rm c}(r)-t]\right\}, \label{sol}
\eeq
where $t_{\rm c}$ is a real-valued arbitrary function, fixed by the initial data via
\beq
t_{\rm c}(r)=\fr{2}{\sqrt{3\Lambda}}\sinh^{-1}\left(\sqrt{\fr{\Lambda r^{3}}{3M}}\right), \label{tc}
\eeq
where the scaling $R(0,r)=r$ was adopted. The relevant derivatives of $R(t,r)$ are:
\bqa
R'(t,r)&=&R\left[\fr{M'}{3M}+\sqrt{\fr{\Lambda}{3}}t'_{\rm c}\coth (t_{\rm c}-t)\right], \label{Rpr} \\
\dot{R}(t,r)&=&-\sqrt{\fr{\Lambda}{3}}R\coth (t_{\rm c}-t),
\eqa
where the minus sign corresponds to implosion.

A complete solution is given by equation (\ref{alph}), where $R$ and $\nu$ are given by equations (\ref{nu})-(\ref{kay}) and (\ref{sol}), together with the energy density 
\beq
\rho=(M'-3M\nu')e^{-\alpha-2\beta}=\fr{M'-3M\nu'}{R^{2}(R'+R\nu')}e^{-2\nu}. \label{rho}
\eeq
Imposing the weak energy condition \cite{hawking&ellis73}, $T_{ab}u^{a}u^{b}\geq0\,\Rightarrow\,\rho>0$, leads to the constraint 
\beq
\nu'\leq\fr{1}{3}\fr{M'}{M}. \label{wec}
\eeq

\begin{figure}
\begin{center}
\epsfxsize=15pc
\epsffile{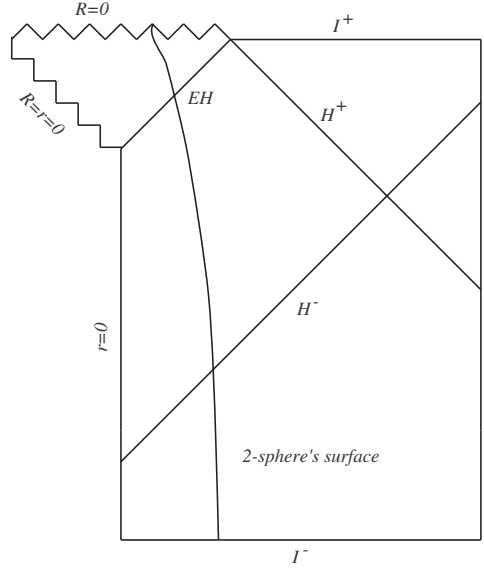}
\end{center}
\caption{Conformal diagram for asymptotically de Sitter quasispherical dust collapse, for the two-surface $\Sigma:\{x,y=\mbox{const.}\}$. The $r=0$ shell is the first one to become singular. The outer shell, which defines the boundary of $\Sigma$, first collapses through the past cosmological horizon ($H^{-}$), and then, after a finite amount of time $t$, through the black hole event horizon (EH). The example depicted here contains a locally naked singularity, whose null portion is visible to local observers inside the EH. Since the EH coincides with the Cauchy horizon, the singularity cannot be globally naked. \label{fig1}}
\end{figure}

\subsection{Regularity conditions}

We assume the following regularity conditions for the spacetime:

(i) There are no shell crossing singularities: $(Re^{\nu})'>0$, which also guarantees that the proper area radius of the shells is a monotonically increasing function of the comoving spatial coordinate $r$,
\beq
R'+R\nu'>0. \label{rc1}
\eeq

(ii) Regularity of the metric at $r=0$ requires that all the $c_{i}(r)$ be at least $C^{1}$ therein, with finite first derivatives, such that
\beq
\lim_{r\rightarrow 0^{+}} c'_{i}(r)<\infty \; \Rightarrow \; \lim_{r\rightarrow0^{+}} \nu'<\infty. \label{rc2}
\eeq

(iii) In order for the metric to be locally Euclidean at $r=0$, we must have
\beq
k(0)=0. \label{rc3}
\eeq

(iv) From a physical point of view, it is reasonable to require that the initial velocity profile, $\dot{R}(0,r)$, and the energy density, $\rho(0,r,x,y)$, be everywhere bounded. With the scaling $R(0,r)=r$, it follows from equation (\ref{szd}) that $M\gtrsim O(r)$ near $r=0$, and equation (\ref{rho}) implies $M'\sim O(r^{2})$; near $r=0$ we must therefore have
\beq
\fr{M'}{M}\sim O(r^{-1}). \label{rc4}
\eeq

\section{Existence and visibility}

Each two-surface $\Sigma:\{t,r=\mbox{const.}\}$ is a two-sphere with proper area radius $R(t,r)e^{\nu(r,x,y)}$, and coordinate center $(\fr{c_{2}}{c_{1}},\fr{c_{3}}{c_{1}})$ on $\Sigma$. Since the center is $r$-dependent, the shells are not in general concentric, which accounts for the absence of spherical (or otherwise) symmetry. Note also that, unlike in the spherically symmetric case, the energy density is not constant over each two-sphere. If three of the $c_{i}$'s are constant, there is a single $c_{i}$ free function---which fixes $k(r)$, via equation (\ref{kay})---and the model reduces to the TBdS case (where the $k=0$ case corresponds to the choice $c_{1}=c_{4}=\fr{1}{2}$ and $c_{2}=c_{3}=0$).

From equation (\ref{rho}), one sees that the energy density diverges at $R=0$, thereby signaling a curvature singularity \cite{szekeres75b}. As in the spherical case, $r>0$ events for which $R(r,t)=0$, are spacelike and hence cannot be naked \cite{szekeres75b,joshi&krolak96,christodoulou84}. Of potential interest is the ``central'' $r=0$ singularity.

In order for this singularity to be at least locally naked, there must exist at least one outgoing non-spacelike geodesic with past endpoint at the singularity. We note that the existence of a solution of the, e.g., outgoing radial null geodesics equation, does not {\em per se} guarantee the local visibility of the singularity. If such geodesics are emitted after the apparent horizon (AH) forms, they will be contained in the causal past of a trapped region (hence unavoidably trapped), and thus any initially diverging geodesic congruence will immediately start reconverging upon emission. Therefore, although there may be outgoing non-spacelike geodesics with past endpoint at the singularity, if they are emitted after the AH forms, they are only defined at the point of emission, and hence we shall not consider them in the operational definition of local nakedness adopted in this paper. We shall show that there are outgoing non-spacelike geodesics that form before or at the time at which the AH does, by analyzing the geodesic behaviour along null and timelike directions.

\subsection{Null geodesics}

The equation for outgoing radial null geodesics (ORNGs) is
\beq
\left(\fr{dt}{dr}\right)_{\rm out}=R'+R\nu'. \label{null}
\eeq
If it admits a regular solution, $t(r)$, in an open neighbourhood containing $r=0$, the singularity is visible---at least locally---provided $t(0)\leq t_{\rm AH}(0)$, where $t_{\rm AH}(r)$ is the AH curve on the $\{t,r\}$ plane. Let us then assume that such a solution exists, and is given to leading order by
\beq
t(r)=t_{0}+ar^{\sigma}, \label{ge}
\eeq
where $a\in\mathbb{R}^{+}$, and $\sigma\in\mathbb{N}^{+}$. Expanding the central energy density near $r=0$ as
\beq
\rho_{\rm c}(r)=\sum^{+\infty}_{i=0} \rho_{i}r^{i},
\eeq
we have, from equations (\ref{rho}), (\ref{wec}), (\ref{rc4}), (\ref{tc}) and (\ref{szd}),
\bqa
M(r)&=&M_{0}r^{3}+M_{n}r^{n+3}+O(r^{n+4}), \label{mass} \\
t_{\rm c}(r)&=&t_{0}+t_{n}r^{n}+O(r^{n+1}), \label{tcoll} \\
R(t,r)&=&\left(\fr{9}{4}\right)^{\fr{1}{3}}\left(M_{0}^{\fr{1}{3}}r+m_{n}r^{n+1}\right) \left(t_{0}+t_{n}r^{n}-t\right)^{\fr{2}{3}}+\;O(r^{n+2+\sigma}), \label{bigr}
\eqa
where $t_{n}$ and $m_{n}$ are real coefficients linear in $M_{n}=(4\pi/n)\rho_{n}$, with $n>0$; $\rho_{n}\equiv(\partial^{n}\rho_{c}/\partial r^{n})_{r=0}$ is the first non-vanishing derivative of the central energy density distribution, and
\beq
t_{0}=\sinh^{-1}\left(\sqrt{\fr{\Lambda}{6M_{0}}}\right). \label{tzero}
\eeq
Since the geodesic must lie on the spacetime, from equations (\ref{ge}) and (\ref{tcoll}) it follows that $\sigma\geq n$. If $\sigma=n$, we have the additional constraint, $a<t_{n}$.

The existence of a self-consistent solution of equation (\ref{null}) is dependent on the $r$-differentiability of $R$ and $\nu$ at $r=0$. From equation (\ref{nu}), we have
\beq
\nu'(r,x,y)=-\fr{c'_{1}(x^{2}+y^{2})+2(c'_{2}x+c'_{3}y)+c'_{4}}{c_{1}(x^{2}+y^{2})+2(c_{2}x+c_{3}y)+c_{4}}.
\eeq
Since all the $c_{i}(r)$ are assumed to be at least $C^{1}$, to leading order their MacLaurin expansion is
\beq
c_{i}(r)=c_{i}^{0}+c_{i}^{n_{i}}r^{n_{i}},
\eeq
where $c_{i}^{n_{i}}\equiv(\partial^{n_{i}} c_{i}/\partial r^{n_{i}})_{r=0}$ is the first non-vanishing derivative of $c_{i}(r)$, and $n_{i}\geq1$. Without loss of generality, we can take $n_{1}=\mbox{min} \{n_{i}, i=1...4\}$, such that near $r=0$ we have
\bqa
\nu'&=&\fr{n_{1}r^{n_{1}-1}}{r^{n_{1}}+\zeta}, \label{nu2} \\
\zeta(x,y)&\equiv&\fr{c_{1}^{0}}{c_{1}^{n_{1}}}+\left(2c_{2}^{0}x+2c_{3}^{0}y+c_{4}^{0}\right)c_{1}^{-n_{1}}(x^{2}+y^{2})^{-1}.
\eqa
If $\zeta=0$ (which is a set of measure zero in the initial data, since it requires $c_{i}^{0}=0$, $i=1...4$), then $\nu'=n_{1}r^{-1}$ and the metric is just $C^{0}$ at $r=0$. In general, $\zeta\neq0$, which renders the metric $C^{1}$ at $r=0$, as desired. In this case, from equations (\ref{null})-(\ref{ge}) and (\ref{bigr}), we obtain, near $r=0$
\bqa
a\sigma r^{\sigma-1}&=&\left(\fr{2n}{3}+1\right)Ar^{\fr{2n}{3}}+\fr{A}{\zeta}n_{1}R^{\fr{2n}{3}+n_{1}}+O(r^{\sigma+1-\fr{n}{3}}), \\
A&\equiv&\left(\fr{9M_{0}}{4}\right)^{\fr{1}{3}}t_{n}^{\fr{1}{3}}.
\eqa
Since $n_{1}\geq1$, the second term on the right-hand-side is to be neglected to leading order in $r$, and a self-consistent solution exists provided
\bqa
a&=&\left(\fr{9M_{0}}{4}\right)^{\fr{1}{3}}t_{n}^{\fr{1}{3}}\left(\fr{2n}{3}+1\right), \\
\sigma&=&1+\fr{2n}{3}. \label{sigma}
\eqa
The condition $\sigma>n$ now reads $n<3$.  For $n=1,2$ (i.e., for $\rho_{1}\neq0$, or $\rho_{1}=0$ and $\rho_{2}\neq0$) there is a self-consistent solution to the ORNG equation in the limit $t\rightarrow t_{0}$, $r\rightarrow0$, and thus there is at least one ORNG starting from the singularity.

Let us now examine the case $n=\sigma=3$. Proceeding as before, we obtain, to leading order,
\beq
3ar^{2}=3\left(\fr{9M_{0}}{4}\right)^{\fr{1}{3}}(t_{3}-a)^{\fr{2}{3}}r^{2},
\eeq
which is identically satisfied provided
\beq
a^{3}-\mu^{3}a^{2}+2\mu^{2}t_{3}a=0, \label{cub}
\eeq
where $\mu\equiv(9M_{0}/4)^{1/3}$. This equation has two non-zero distinct roots (other than the $a=0$ trivial root), given by $a=(\mu^{2}/2)\pm\sqrt{\mu^{4}-8t_{3}}$, if $t_{3}<\fr{1}{8}\left(\fr{9M_{0}}{4}\right)^{\fr{4}{3}}$, which imposes a constraint on $\rho_{3}$, for a given $\rho_{0}$. In addition, one must also require $a<t_{3}$, which leads to
\beq
\fr{\mu^{2}}{2}-4+\sqrt{\mu^{4}+16-4\mu^{2}}<t_{3}<\fr{1}{8}\mu^{4}. \label{t3c}
\eeq
Hence, as long as one restricts ourselves to initial data that satisfies the above condition, there exists an ORNG with past endpoint at the singularity. We note, however, that the $n=3$ case is less generic than the $n<3$ cases, as it requires that $\rho_{1}=\rho_{2}=0$ and $\rho_{3}$ obey condition (\ref{t3c}).

Summarizing, for a given initial density profile, the ORNG equation admits a regular solution at $r=0$, independently of the details of $\nu(r,x,y)$ (provided $\nu$ obeys the regularity condition (\ref{rc2})).

\subsection{Timelike geodesics}

Let us consider radial timelike geodesics (RTGs) described by a tangent vector field $K^{a}=\fr{dx^{a}}{d\tau}$, where $\tau$ is an affine parameter along the geodesic. A sufficient set of equations for $K^{a}$ is
\bqa
&&K^{t}=\pm\sqrt{1+e^{2\alpha}(K^{r})^{2}}, \\
&&(K^{r})\dot{}+\fr{K^{r}}{K^{t}}(K^{r})'R'+K^{r}\fr{\dot{R}'+\dot{R}\nu'}{R'+R\nu'}+\fr{(K^{r})^{2}}{K^{t}}\fr{R''+R'\nu'+R\nu''}{R'+R\nu'}=0, \label{ge3}
\eqa
where the first equation is the unit-norm condition, and the second one follows from the geodesic equation. By inspection, one sees that the above set admits the trivial solution
\bqa
K^{t}&=&\pm1, \label{kt}\\
K^{r}&=&0, \label{kr}
\eqa
which leads to
\bqa
t=t_{0}\pm(\tau-\tau_{0}), \label{timet} \\
r=r_{0}=\mbox{const.},
\eqa
where $\tau_{0}$ is the proper time at which RTG's depart (arrive at) the central singularity, and the plus or minus sign refers to outgoing or ingoing RTGs, respectively. The outgoing RTG departing from the singularity is given
by $r=0$, and $t=t_{0}+\tau-\tau_{0}$, and thus does not
belong in the spacetime. The ingoing RTG is given by $r=0$,
$t=t_{0}-\tau+\tau_{0}$, where $t_{0}=t_{\rm c}(0)=0$ is the time at which the RTG arrives at the singularity.  

Since equation (\ref{ge3}) is a mixed first-order linear PDE for $K^{r}(t,r)$, its solution need not be unique \cite{kevorkian99}. Indeed, one can explicitly construct other families of solutions, as follows. Near the singularity, we can write, to leading order
\bqa
t_{\rm RTG}(r)&=&t_{0}+br^{p}, \label{rtg} \\
R(t_{\rm RTG}(r),r)&=&a_{0}r^{\fr{2n}{3}+1}+O(r^{p+2-\fr{n}{3}}), \\
R'(t_{\rm RTG}(r),r)&=&a_{0}(\fr{2n}{3}+1)r^{2n/3}+O(r^{p+1-\fr{n}{3}}),
\eqa
where $a_{0}\in{\mathbb R}^{+}$, $b\in{\mathbb R}$, and $p\geq n$ (such that the geodesics thus constructed belong in the spacetime).

Let us now assume that $K^{r}(t,r)\propto (t-t_{0})^{\alpha}r^{\beta}$, where $\alpha,\beta\in {\mathbb R}$. From equation (\ref{rtg}), along the RTG, we have then
\beq
K^{r}(t_{\rm RTG}(r),r)=kr^{\alpha p+\beta}. \label{kanstz}
\eeq
where $k\in{\mathbb R}$ is a constant, yet to be determined. With these ansatze, the RTG equation reads
\beq
\fr{dt_{\rm RTG}}{dr}=bpr^{p-1}=\fr{K^{t}}{K^{r}}=\sqrt{(K^{r})^{-2}+(R'+R\nu')^{2}}. \label{rtge}
\eeq
Now, from equation (\ref{nu2}), we have $\nu'=n_{1}\zeta^{-1}r^{n_{1}-1}$, where $n_{1}\geq1$ is fixed by the initial data, and characterizes the differentiability of the metric at the origin. From equation (\ref{rtge}), it follows that
\beq
b^{2}p^{2}r^{2(p-1)}=k^{-2}r^{-2(\alpha p+\beta)}+a_{0}^{2}q^{2}r^{2l}+a_{0}^{2}\fr{n_{1}^{2}}{\zeta^{2}}r^{2(q+n_{1}-1)}+2a^{2}_{0}q\fr{n_{1}}{\zeta}r^{q+l+n_{1}-1}, \label{eqx}
\eeq
where $q=1+2n/3$, and $l=2n/3$. Now, since $n_{1}\geq1$, a straightforward calculation shows that if $n<6$, the last two terms are of higher order than the remaining ones. We shall assume this for now, and conclude below that a self-consistent solution always satisfies this condition. We then look for solutions which are homogeneous in the leading order of $r$. This leads to a coupled algebraic system for the parameters $p,q,\alpha,\beta$:
\bqa
p-1=-\alpha p-\beta, \\
\alpha p+\beta=-q+1,
\eqa
which is solved by
\beq
p=1+\fr{2n}{3}=\fr{1-\beta}{1+\alpha}. \label{solu}
\eeq
The constraint $p\geq n$ now reads $n\leq3$, which is consistent with the earlier assumption $n<6$. For parameter values satisfying equation (\ref{solu}), equation (\ref{rtge}) becomes
\beq
C(a_{0},b,p,k)r^{4n/3}=0,
\eeq
whose solution is given by the algebraic constraint
\beq
C(a_{0},b,p,k)\equiv b^{2}p^{2}-k^{-2}-a_{0}^{2}\fr{4n^{2}}{9}=0. \label{constr}
\eeq
Now, since $K^{r}$ is obtained from the derivatives of $R(t,r)$, the parameter $k$ is not independent of $a_{0}$; therefore, for given initial data, the set of parameters $\{a_{0},p,k\}$ is uniquely determined. From equation (\ref{constr}), it then follows that $b$ is entirely determined from the initial data, modulo its sign, i.e., for a a given set of initial data, both ingoing ($b<0$) and outgoing ($b>0$) radial timelike geodesics exist.

\subsection{Local visibility}

Thus far we have shown the existence of outgoing null and timelike geodesics with past endpoint at the singularity. In order for these geodesics to be visible to local non-spacelike observers, they must not be contained in the domain of dependence of any trapped surfaces. Clearly, a necessary and sufficient condition for the singularity to be locally naked is that future-directed geodesics departing from the singularity, do so before or at the time at which the AH forms; otherwise they will be unavoidably trapped. 

The AH is the outer boundary of a trapped surface---a compact spacelike two-surface whose outgoing and ingoing null geodesic congruences have vanishing expansion. Let us then consider the two-surface $\Sigma: \{t,r=\mbox{const}\}$, and a congruence of radial null geodesics, with tangent vector field $\xi^{a}=\fr{dx^{a}}{d\lambda}$ (where $\lambda$ is an affine parameter along the geodesic), orthogonal to it. The expansion of such a congruence is given by the scalar
\beq
\Theta\equiv\nabla_{a}\xi^{a}.
\eeq
Orthogonality to $\Sigma$ allows the choice
\beq
\xi^{x}=\xi^{y}=0,
\eeq
and the null-norm condition gives
\beq
\left(\xi^{t}\right)^{2}=(R'+R\nu')^{2}\left(\xi^{r}\right)^{2}.
\eeq
One can choose the affine parameter such that
\bqa
\xi^{t}&=&(R'+R\nu')^{2}, \label{xit} \\
\xi^{r}&=&\epsilon=\pm1, \label{xir}
\eqa
where the plus or minus sign corresponds to outgoing or ingoing geodesics, respectively. From equations (\ref{szm})-(\ref{kay}) and (\ref{xit})-(\ref{xir}), we have then
\beq
\Theta=\fr{2}{R}(R'+R\nu')(\dot{R}+\epsilon).
\eeq
Now, from equation (\ref{rc1}) the first term is always positive-definite, and thus $\mbox{sign}\,\Theta=\mbox{sign}\,(\dot{R}+\epsilon)$. Since we are interested in implosion situations, we must also have $\dot{R}\leq0$ for all times. It then follows that for ingoing null geodesics ($\epsilon=-1$), we have $\Theta<0$, i.e., their expansion is always negative (convergence). For outgoing geodesics ($\epsilon=+1$) one can have initial divergence, if $\dot{R}(0,r)>-1$. However, since\footnote{Near $r=0$ this condition becomes $\rho_{c}(0)\gtrsim\Lambda$, which is always satisfied for physically reasonable values of the central density and cosmological constant (in geometrized units).} $\ddot{R}(0,r)=-\fr{M(r)}{r^{2}}+\fr{\Lambda}{3}r<0$, there is a finite time $t_{\rm M}(r)>0$ such that $\dot{R}(t_{\rm M},r)<-1$. Therefore, there is a time $t_{\rm AH}\in(0,t_{\rm M})$, such that 
\beq
\dot{R}(t_{\rm AH},r)=-1.
\eeq
It then follows from equation (\ref{szd}) that the AH is given by the curve $t_{\rm AH}(r)$, where $R(t_{\rm AH},r)$ is a solution of
\beq
\fr{\Lambda}{3}R^{3}-R+2M=0.
\eeq
This equation has three distinct real roots if $3M\sqrt{\Lambda}<1$, two of which are positive and given by
\beq
R_{\pm}=\fr{2}{\sqrt{\Lambda}}\sin\left[\fr{1}{3}\sin^{-1}(3M\sqrt{\Lambda})+\fr{2\pi}{3}\delta(1\pm1)\right], \label{rbh}
\eeq
where $\delta(x)$ is the Dirac delta-function, and $R_{-}>R_{+}>0$, corresponding to the choice $0\leq\sin^{-1}\omega\leq\pi/2$, $0\leq\omega\leq1$. The third root, $R_{3}=-R_{-}-R_{+}$, is negative and hence unphysical. $R_{-}$ is a generalized cosmological horizon ($R_{-}=\sqrt{3/\Lambda}$, when $M=0$) and $R_{+}$ the black hole apparent horizon ($R_{+}=2M$ when $\Lambda=0$; the apparent and event horizons coincide in the static case). For $3M\sqrt{\Lambda}=1$, the two horizons coincide. If $3M\sqrt{\Lambda}>1$, there is one negative real root and two complex (conjugate) roots, all of which are unphysical: the spacetime does not admit any horizons in this case. From the ``+'' solution in equation (\ref{rbh}), together with equations (\ref{sol})-(\ref{tc}), we obtain
\beq
t_{\rm AH}(r)=t_{\rm c}(r)-\varsigma(r), \label{tah}
\eeq
where
\beq
\varsigma(r)\equiv\fr{2}{\sqrt{3\Lambda}}\sinh^{-1}\left\{\left(\fr{8}{3M\sqrt{\Lambda}}\right)^{1/3}\left[ \sin\left(\fr{1}{3}\sin^{-1}(3M\sqrt{\Lambda})\right)\right]^{3/2}\right\}. \label{xior}
\eeq
At the origin, $\varsigma(0)=0$, and thus $t_{\rm AH}(0)=t_{\rm c}(0)$. Since the AH and the singularity curve form at the same time at $r=0$, the visibility of the singularity is determined by the relative slopes of the curves $t_{\rm AH}(r)$ and $t_{\rm ORG}(r)$ on the $\{t,r\}$ plane (where `ORG'  denotes non-spacelike outgoing radial geodesics). The singularity is (at least) locally naked iff 
\beq
\lim_{r\rightarrow0}\left(\fr{dt_{\rm AH}}{dr}\right)/\left(\fr{dt_{\rm ORG}}{dr}\right)>1. \label{vis}
\eeq

For ORNG's, given by $n<3$, from equations (\ref{ge}) and (\ref{sigma}), we get
\beq
\lim_{r\rightarrow0}\left(\fr{dt_{\rm AH}}{dr}\right)/\left(\fr{dt_{\rm ORNG}}{dr}\right)=\lim_{r\rightarrow0} \fr{nt_{n}}{a\sigma}r^{-1+n/3}=+\infty,
\eeq
and hence the singularity is locally naked. This is in agreement with the analysis of \cite{deshingkaretal00}, which shows that the value of $\Lambda$ does not alter the visibility of the singularity for $n<3$. (For the highly non-generic case $n=3$, part---but {\em not} all---of the singularity spectrum may be covered for sufficiently large (but finite) $\Lambda$; the method discussed in the present paper is not appropriate for such an analysis, and we thus refer to reader to reference \cite{deshingkaretal00} for further details).

For ORTG's, given by equation (\ref{rtge}), the local visibility condition reads
\beq
\lim_{r\rightarrow0} \fr{nt_{n}}{bp}r^{n-p}>1.
\eeq
Clearly, we must have $p>\sigma>n$, which yields $\lim_{r\rightarrow0} \fr{nt_{n}}{bp}r^{n-p}=+\infty$. Hence, the singularity is locally visible along such ORTG's.
\begin{figure}
\begin{center}
\epsfxsize=20pc
\epsffile{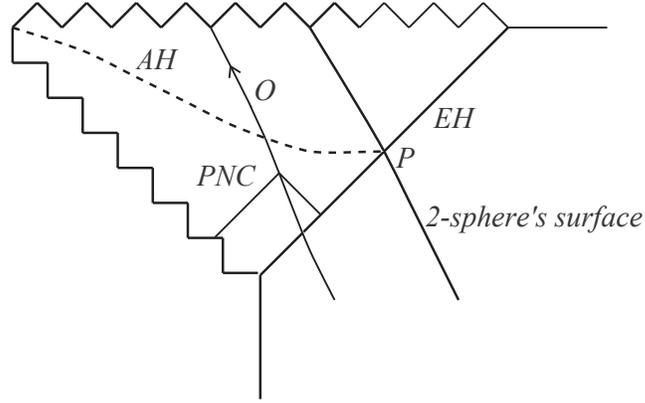}
\end{center}
\caption{Locally naked singularity in quasispherical asymptotically de Sitter dust collapse. The dashed line is the apparent horizon (AH), which joins the event horizon (EH) when the boundary of the two-surface $\Sigma$ crosses the latter, at event $P$. Also shown is the worldline of a timelike observer $O$, who penetrates the EH and terminates at the spacelike portion of the singularity. PNC denotes the past null cone of $O$, which intersects the EH {\em and} the null portion of the singularity: clearly, there are non-spacelike geodesics emitted from the singularity, before the AH forms, which are contained in the PNC of $O$; the singularity is therefore locally naked. \label{fig2}}
\end{figure}

\section{Curvature strength}

Let $\gamma(\tau)$ be a RTG, with tangent vector $K^{a}=\fr{dx^{a}}{d\tau}$, that terminates at the singularity at $\tau=\tau_{0}$. Such a singularity is said to be Tipler strong along $\gamma(\tau)$ if the volume three-form $V(\tau)$ vanishes in the limit $\tau\rightarrow\tau_{0}$ \cite{tipler77}. If the scalar $\Psi\equiv R_{ab}K^{a}K^{b}$ obeys the {\em strong limiting focusing condition},
\beq
\lim_{\tau\rightarrow\tau_{0}} (\tau-\tau_{0})^{2}\Psi>0,
\eeq
then the singularity is gravitationally strong in the sense of Tipler \cite{clarke&krolak85}. This sufficient condition guarantees that any three-form defined along $\gamma(\tau)$ vanishes in the limit $\tau\rightarrow\tau_{0}$, due to unbounded curvature growth.

From equations (\ref{szm})-(\ref{kay}) and (\ref{kt})-(\ref{kr}), we have
\bqa
\Psi&=&(1+\xi)^{-1}\left(\fr{M'}{R^{2}R'}+\fr{4M}{R^{3}}\xi\right)-\Lambda, \\
\xi&\equiv&\fr{R}{R'}\nu'.
\eqa
Now, from equation (\ref{Rpr}) it follows that $\lim_{r\rightarrow0^{+}} \fr{R}{R'}=0$, and from equation (\ref{wec}) we have $\lim_{r\rightarrow0^{+}} \nu'<\infty$. Therefore, $\lim_{r\rightarrow0^{+}} \xi=0$, which implies
\beq
\lim_{r\rightarrow0^{+}} \Psi=\lim_{r\rightarrow0^{+}} \fr{M'}{R^{2}R'}-\Lambda.
\eeq
Using equations (\ref{mass})-(\ref{bigr}), together with equation (\ref{timet}), we obtain
\bqa
\lim_{r\rightarrow0^{+}} \Psi&=&\fr{2}{3}\eta^{-\fr{4}{3}}\left(\eta^{\fr{2}{3}}+\fr{2}{3}nt_{n}r^{n}\eta^{-\fr{1}{3}}\right)^{-1}, \\
\eta&\equiv&t_{n}r^{n}\mp(\tau-\tau_{0}).
\eqa
Hence, at $r=0$ we have
\beq
\lim_{\tau\rightarrow\tau_{0}} (\tau-\tau_{0})^{2}\Psi=\fr{2}{3},
\eeq
and the singularity is therefore Tipler strong. That is, along the timelike geodesics given by equations (\ref{kt})-(\ref{kr}), the singularity is Tipler strong, independently of the details of the initial density distribution. We note that, since all the sufficient criteria for Tipler strong singularities are sufficient for  {\em deformationally strong} ones \cite{ori00}, the central curvature singularity in TBdS collapse is also deformationally strong.

\section{Concluding remarks}

We found that there are initial data, for which null and timelike families of outgoing radial geodesics with past endpoint at the singularity exist, which are emitted along the apparent horizon curve; the singularity is therefore at least locally naked. When examined along timelike geodesics terminating at the singularity, the latter was found to be Tipler strong, regardless of the initial density distribution. This constitutes a rather robust result, in that it holds true independently of the initial data [the initial density profile $\rho(0,r,x,y)$, for any given velocity profile $\dot{R}(0,r)<0$], thereby implying stability against perturbations of the latter.

It was shown in \cite{deshingkar&jhingan&joshi98} that there is no directional dependence of the local visibility in asymptotically flat Szekeres dust collapse. We obtained the same result for asymptotically de Sitter quasispherical dust collapse, as expected. We also showed the absence of effects of $\Lambda$ on the visibility of singularities for $n<3$, already observed in spherical Tolman-Bondi-de Sitter collapse. The fact that our results are qualitatively equivalent to those for spherical inhomogeneous collapse (with or without a cosmological constant), suggests that neither the lack of asymptotic flatness, nor mild (in the well-defined geometric sense of Szekeres metrics) departures from spherical symmetry change the standard picture of singularity formation and structure in gravitational collapse. 

Regarding the generality of the present results insofar as the matter content is concerned, two points are worth mentioning. Firstly,we considered the special $k(r)=0$ case (cf. equation (\ref{kay})), which corresponds to gravitationaly unbound matter configurations. This simplifying assumption ($k=0$) was found not to qualitatively change the corresponding results for spherical LTB collapse \cite{jhingan&joshi97} (note, however, that in the asymptotically flat case, $k=0$ corresponds to marginally bound matter configurations). Since the structure of the central singularity in the present case is analogous to that of LTB collapse (the dependence of the visibility of the singularity on the initial data is exactly the same), we do not expect the inclusion of $k(r)\neq0$ to qualitatively change the end result of collapse. Secondly, a more general issue---that concerns not only the present analysis, but the issue of generic gravitational collapse---is that of the physical reasonability of ``dust''. Whilst early studies of high-density nuclear matter suggested that such an approximation could be a legitimate one  (at least for spherical collapse) \cite{hagedorn65}, it is now well-known that radial and tangential pressures {\em must} be taken into account at the very late stages of realistic astrophysical collapse \cite{miller&sciama80}. The inclusion of radial pressure in spherical dust collapse has been recently considered \cite{goncalves&jhingan01}, and it has been found to cover part---but not all---of the singularity spectrum, i.e., configurations that would otherwise develop locally naked singularities, end up in a black hole. The inclusion of tangential pressure has been considered for special cases of the Einstein cluster class \cite{hin1,jhingan&magli00} and for more generic configurations \cite{goncalves&jhingan&magli01}, with markedly different results: tangential stresses tend to uncover part of the singularity spectrum, i.e., configurations that would otherwise terminate in a black hole, develop a locally naked singularity. For the reasons mentioned above, we expect the inclusion of radial and tangential stresses in quasispherical asymptotically de Sitter dust collapse to produce similar effects to those of the spherical case.

Unlike the matter content, however, the effects of non-sphericity on the endstate of collapse are much less clear, and further analyses, of stronger deviations from spherical symmetry, need to be undertaken to confidently establish the role of the latter in the final state of generic gravitational collapse.

\ack

It is a pleasure to thank J. Barrow and K. Thorne for helpful comments and discussions, and F. Mena and B. Nolan for privately communicating their recent results on geodesics in spherical collapse \cite{mena&nolan01}, which led to a correction in this manuscript. This work was supported by FCT (Portugal) Grant PRAXIS XXI-BPD-16301-98, NSF Grant AST-9731698, and by NSF Grant PHY-0099568.

\end{document}